\documentclass[epj,final]{svjour}
\usepackage{graphics}
\usepackage{graphicx}
\usepackage{amssymb}

\begin{document}

\title{Phase transitions in two-dimensional anisotropic quantum magnets}
\author{Alessandro Cuccoli\inst{1} \and  Tommaso Roscilde\inst{2} \and  Valerio
  Tognetti\inst{1} \and Ruggero Vaia\inst{3} \and Paola
  Verrucchi\inst{1} }
\institute{Dipartimento
  di Fisica dell'Universit\`a di Firenze and Istituto Nazionale di
  Fisica della Materia (INFM), \\ Largo E. Fermi~2, I-50125 Firenze,
  Italy \and
  Dipartimento di Fisica ``A. Volta'' dell'Universit\`a di Pavia and 
  Istituto Nazionale di Fisica della Materia (INFM),\\
  via Bassi~6, I-27100 Pavia, Italy  \and  
  Istituto di Elettronica Quantistica
    del Consiglio Nazionale delle Ricerche,
    via Panciatichi~56/30, I-50127 Firenze, Italy,
    \\ and Istituto Nazionale di Fisica della Materia (INFM)
} % end institute

\date{\today}

\abstract{
We consider quantum Heisenberg ferro- and antiferromagnets on the square
 lattice with exchange anisotropy of easy-plane or easy-axis
type. The thermodynamics and the critical behaviour of the
models are studied by the pure-quantum self-consistent harmonic
approximation, in order to evaluate the spin and anisotropy dependence of
the critical temperatures. Results for thermodynamic quantities are 
reported and comparison with experimental and numerical
simulation data is made. The obtained results allow us to 
draw a general picture of the subject and, in particular, to estimate the
value of the critical temperature for any model belonging
to the considered class. 
\PACS{
{75.10.Jm} {Quantized spin models}\and
{75.10.-b} {General theory and models of magnetic ordering}\and
{75.40.Cx} {Static properties}
   } % end of PACS codes
} %end of abstract

\maketitle

\section{Introduction}
\label{s.intro}

In this paper we examine the class of magnetic models described by the
Hamiltonian
\begin{equation}
\hat{\cal{H}}= -J\sum_{<{\bf i j}>}~\left[
\mu\left(\hat{S}^x_{\bf i}\hat{S}^x_{\bf j}+
         \hat{S}^y_{\bf i}\hat{S}^y_{\bf j}\right)+
\lambda\hat{S}^z_{\bf i}\hat{S}^z_{\bf j}\right]
\label{e.H}
\end{equation}
where ${\bf i}\equiv(i_1,i_2)$, ${\bf j}\equiv(j_1,j_2)$ are sites on a
square lattice, and the sum
runs over the pairs of nearest neighbours; the spin degrees of freedom
are described by the quantum operators 
$\hat{\bf S}_{\bf i}\equiv 
(\hat{S}^x_{\bf i},\hat{S}^y_{\bf i},\hat{S}^z_{\bf i})$, obeying the
angular momentum commutation relations 
$[\hat{S}^\alpha_{\bf i},\hat{S}^\beta_{\bf j}]=
\epsilon^{\alpha\beta\gamma}\delta_{\bf ij}\hat{S}_{\bf i}^\gamma$
with $|\hat{\bf S}|^2=S(S+1)$.

For positive $\lambda$ and $\mu$, the sign of the exchange integral $J$
sets the ferro- or antiferromagnetic
character of the model, $J>0$ or $J<0$ respectively; for the sake of a
more compact notation, however, we will hereafter take a positive $J$ and
properly change the sign of $\mu$ and $\lambda$ to obtain the required 
antiferromagnetic Hamiltonian.

At a classical level, i.e. when  $S\to \infty$, ferro- and
antiferromagnets have the same static properties and can be
simultaneously studied as far as their thermodynamics is concerned.
On the other hand, when quantum fluctuations are taken into account, the
two types of models display substantially different features.
It is worthwhile noticing that, despite displaying the most interesting
quantum features, antiferromagnetic models are less affected
by quantum fluctuations than ferromagnetic ones, as quantum
renormalizations are sensibly stronger in the latter, whatever the 
approach used to evaluate them.

In this paper, the anisotropy parameters $\mu$ and $\lambda$ are set
to range in the interval $[-1,1]$, and the following sub-classes of models
are hence considered:

(i) Isotropic: $\mu=\lambda$, $|\lambda|=1$,  ferro- and
antiferromagnetic ($\lambda>0$ and $\lambda<0$, respectively).

(ii) Easy-plane: $\mu=1$, $\lambda\in (-1,1)$, ferro- and
antiferromagnetic ($\lambda>0$ and $\lambda<0$, respectively).

(iii) Easy-axis: $\mu\in (-1,1)$, $\lambda={\rm sign}(\mu)$,
ferro- and antiferromagnetic ($\mu>0$ and $\mu<0$, respectively).

Models belonging to these sub-classes have the common feature of
displaying, 
as the temperature decreases, a phase-transition towards a more ordered
phase; this transition generically occurs at a critical temperature which
is a function of both the spin value and the anisotropy parameters,
and will be hereafter indicated by $t_{\rm c}(\nu,S)$, with 
$\nu=\mu,\lambda$ depending on the specific class considered, and
the reduced temperature $t$ defined by $t\equiv T/J\widetilde{S}^2$ with
$\widetilde{S}\equiv S+1/2$.

(i) The isotropic model, both ferro- and antiferromagnetic, has the
full rotational 
$O(3)$ symmetry and it is hence in a paramagnetic phase down to $t=0$, as required by
the Mermin-Wagner theorem~\cite{MerminW66}. Its ground state, however and
at variance with the one-dimensional case, is ordered for all
spin values also in the antiferromagnetic case~\cite{Manousakis91},
and the model can hence be thought to have a phase transition at a
critical temperature $t_{\rm c}(\pm 1,S)=0$ $\forall S$.

(ii) In the easy plane case, the 
Mermin-Wagner theorem still holds and no finite temperature transition
towards a phase with a finite order parameter, may occur. 
However, a Berezinskii-Kosterlitz-Thouless (BKT) transition
~\cite{Berezinskii70,KosterlitzT73}, related with
the existence of vortex-like topological excitations,  is known to
characterize the class of the easy-plane models and may
occur at a critical temperature $t_{\rm BKT}(\lambda,S)>0$.

The reference system for the easy-plane class is the {\it planar}, or
$XY$, model, defined by Eq.(\ref{e.H}) with $\lambda=0$,
$\mu=1$. Its BKT critical temperature  has been seen to be finite for all 
spin values: above $t_{\rm BKT}$ the system is disordered, with
exponentially decaying correlation functions. In the whole
region $0<t<t_{\rm BKT}$ the system is in a critical phase with vanishing 
magnetization: its correlation functions decay according to a power law,
testifying to the existence of {\it quasi}-long-range
order; at $t=0$ the magnetization gets a finite value and the system is
ordered. The $XY$ model may be thought to display two phase-transitions:
a BKT one at a finite temperature, followed by one towards a fully ordered
phase at $t=0$. 

(iii) In the easy-axis case, the Mermin-Wagner theorem does not hold,
and the easy-axis models may undergo a transition of the Ising (I)
type towards a phase with long-range order, at a critical temperature
$t_{\rm I}(\mu,S)>0$. 

The reference system for this
class of models will be hereafter called the $Z$ model, described by
Eq.(\ref{e.H}) with $\mu=0$, $\lambda=1$; such model has a transition
temperature
$t_{\rm I}(0,S)>0$ for all spin values. Above $t_{\rm I}$ the system
is
in
a paramagnetic phase, with exponentially decaying correlation functions;
below $t_{\rm I}$ the system is ordered with a finite magnetization
along the easy axis. The $Z$ model with $S=1/2$ is the Ising
model, representing an important source of information for the whole
class of the easy-axis models, not only because of the existing exact
results by Onsager~\cite{Onsager44}, but also because of its fundamental
role in the renormalization-group approach and in  conformal field theory.

The information on the reference models, as well as the general
considerations based on the symmetries of the Hamiltonian, are surely
valuable, but not sufficient to fully appreciate the richness of
features contained in Eq.(\ref{e.H}), and to properly understand the
thermodynamic and critical behaviour of the many existing real compounds
whose magnetic behaviour is described by such Hamiltonian.
To accomplish this goal one has to study how the specific values of
$\lambda$, $\mu$, and $S$ may determine the behaviour of the system.

At a qualitative level, we know that quantum fluctuations, whose strength
is measured by the quantum coupling $1/S$, introduce a disordering agent in
the classical ($S\to\infty$) picture: as a consequence, we expect the
possible critical temperatures to decrease with decreasing $S$.
The same effect is obtained by increasing $|\lambda|$ (easy-plane case), 
or $|\mu|$ (easy-axis case): this in fact means to weaken the anisotropic
character of the
system, thus allowing larger spin fluctuations out of the
easy plane or the easy axis, and inducing a lowering of the critical
temperature. To this respect it is to be noticed that most real compounds
have very weak anisotropies~\cite{ArtsDW90}, i.e. values of $\lambda$ or
$\mu$ very close
to unity: this is the region where quantitative predictions are more
difficult to obtain, as the model's character is not well defined
and fluctuations are large.

Aim of this paper is to provide a quantitative description of the model 
(\ref{e.H}) when $S$ and $\lambda$, or $\mu$, are varied. We will
mainly concentrate on the values of the critical temperatures, but also
give expressions and results for other thermodynamic quantities. 
The method used is the pure-quantum
self-consistent harmonic approximation (PQSCHA), extensively described in
Ref.~\cite{CTVV92} and already applied to many magnetic
systems (see for instance Refs.~\cite{CTVVprb92,CTVV97}). 
Its implementation is markedly different depending
on the specific class of models considered, as explained in Section
\ref{s.PQSCHA}, and the easy-plane and easy-axis cases will be hence considered
separately (in Section \ref{s.easypl} and \ref{s.easyax},
respectively).
In Section \ref{s.global} we will comment on the resulting global picture 
 and compare our results with all the available
experimental and numerical simulation data. Conclusions about the
critical behaviour of the models will be finally drawn in Section
{\ref{s.concl}.

\section{Method and spin-boson transformations}
\label{s.PQSCHA}

The PQSCHA is a semiclassical method based on the path-integral
formulation of quantum statistical mechanics: the main peculiarity of
the method is that
of defining an analytical separation between classical and 
pure-quantum contributions to the thermodynamics of the system
~\cite{CGTVV95}, thus
allowing the possible exact treatment of the former, while requiring a
self-consistent harmonic approximation of the latter. 
This means that the only approximated contribution is the non-linear
pure-quantum one, which is in fact considered at the one-loop level. The
value of such result is evident if one considers that most of the peculiar
features of magnetic systems, including their critical behaviour, are
determined by long-wavelength excitations, whose character is essentially
classical; the possibility to fully take into account their
non-linearity at the classical level, together with the fact that harmonic
pure-quantum fluctuations are exactly considered, justifies the
success of the PQSCHA
in predicting most thermodynamic properties of magnetic systems in a large
temperature range. 

Without going into the details of the derivation of the PQSCHA, we report
here the essential formulas, and discuss the scheme of implementation to 
magnetic systems, where differences between the sub-classes of models 
become essential.

Consider a quantum spin system on a lattice described by the Hamiltonian
$\hat{\cal H}(\underline{\hat{\bf S}})$ with 
$\underline{\hat{\bf S}}\equiv (\hat{\bf S}_1,...\hat{\bf S}_N)$ and $N$
number of lattice sites; if $\hat{O}$ is the quantum operator
relative to a physical observable of the system, the PQSCHA
expression of its quantum statistical average has the clas\-sical-like form
\begin{equation}
\langle\,\hat{O}\,\rangle={1\over {\cal Z}_{\rm eff}}
\int d^{\scriptscriptstyle N}\!{\bf s}~O_{\rm eff}(\underline{\bf s})~
e^{\,\textstyle{-\beta{\cal H}_{\rm eff}(\underline{\bf s})}}
\label{e.aveO}
\end{equation}
where $\beta=t^{-1}$,
${\cal Z}_{\rm eff}=\int d^{\scriptscriptstyle N}\!{\bf s}\,
e^{\,\textstyle{-\beta{\cal H}_{\rm eff}}}$, and 
$\underline{\bf s}\equiv ({\bf s}_1,...{\bf s}_N)$ with ${\bf s}_i$
classical unit vectors. The effective Hamiltonian 
${\cal H}_{\rm eff}(\underline{\bf s};t,S)$, is a
classical-like function, depending on the parameters of the original
Hamiltonian, and determined by the PQSCHA renormalization procedure;
a similar procedure, when applied to the quantum operator $\hat{O}$,
leads to the function ${\cal O}_{\rm eff}(\underline{\bf s};t,S)$

The phase-space integral in Eq.~(\ref{e.aveO}) may be evaluated by any
classical technique, such as the classical Monte Carlo (MC)
simulation. Eq.~(\ref{e.aveO}), on the other hand, contains quantum
renormalizations embodied in the temperature and spin dependence of 
${\cal H}_{\rm eff}$ and ${\cal O}_{\rm eff}$; such dependence causes the
effective classical model to change for each temperature
point, so that {\it ad hoc} simulations must be performed, and only in
very peculiar cases one can directly use existing
MC data (see for instance the case of the isotropic model in 
Ref.~\cite{CTVV97}).

The PQSCHA naturally applies to bosonic systems, whose Hamiltonian is
written in terms of conjugate operators 
$\underline{\hat{q}}\equiv (\hat{q}_1,...\hat{q}_N)$, 
$\underline{\hat{p}}\equiv (\hat{p}_1,...\hat{p}_N)$ 
such that $[\hat{q}_{m},\hat{p}_{n}]=i\delta_{mn}$; 
the method, however, does not require $\hat{\cal
H}(\underline{\hat{p}},\underline{\hat{q}})$ to be standard, i.e. with
separate quadratic kinetic $\underline{\hat p}$-dependent and  potential
$\underline{\hat q}$-dependent terms, and  
its application may be extended also to magnetic systems,
according to the following scheme~\cite{CGTVV95}:
The spin Hamiltonian $\hat{\cal H}(\underline{\bf S})$ is mapped to
$\hat{\cal H}(\underline{\hat{p}},\underline{\hat{q}})$ by a suitable
spin-boson transformation; 
once the corresponding Weyl symbol ${\cal H}(\underline{p},\underline{q})$,
with $\underline{p}\equiv (p_1,...p_N)$ and 
$\underline{q}\equiv (q_1,...q_N)$ classical phase-space variables,
has been determined~\cite{Weyl50}, the PQSCHA renormalizations may be
evaluated and 
${\cal H}_{\rm eff}(\underline{p},\underline{q})$ and
${\cal O}_{\rm eff}(\underline{p},\underline{q})$ follow. 
Finally ${\cal H}_{\rm eff}(\underline{\bf s})$ and 
${\cal O}_{\rm eff}(\underline{\bf s})$ are
constructed by the inverse of the classical analogue of the
spin-boson transformation used at the beginning.

In order to succesfully carry out such renormalization scheme, the Weyl
symbol of the bosonic Hamiltonian must be a well-behaved function in the
whole phase space. Spin-boson transformations, on the other hand, can
introduce singularities as a consequence of the
topological impossibility of a global mapping of a spherical phase space
into a flat one. 
The choice of the transformation must then be such that the
singularities occur for configurations which are not
thermodynamically relevant, and whose contribution may be hence 
approximated.
Most of the methods for studying magnetic systems do in fact share this
problem with the PQSCHA; what makes the difference is that by using the
PQSCHA one separates the classical from the pure-quantum contribution to
the thermal fluctuations, and the approximation only regards the latter,
being the former exactly taken into account when the
effective Hamiltonian is cast in the form of a classical spin Hamiltonian.

The spin-boson transformation which constitutes the first step of the
magnetic PQSCHA  is chosen  according to the symmetry properties of the
original Hamiltonian and of its ground state.

In markedly easy-plane cases ($|\lambda|\ll 1$) we use, for each spin
operator, the Villain transformation (VT)~\cite{Villain74}
\begin{equation}
\hat{S}^+=e^{i\hat{q}}
          \sqrt{\widetilde{S}^2-\left(\hat{p}+{1\over 2}\right)^2}~,~~
\hat{S}^-=(\hat{S}^+)^\dagger~,~~\hat{S}^z=\hat{p}~;
\label{e.Villain}
\end{equation} 
this transformation keeps the $O(2)$ symmetry in the easy plane, 
meanwhile allowing to deal with the square root in
terms of a physically sound small-$\hat{p}$ approximation.

In the easy-axis case, it makes no sense to use
Eqs.(\ref{e.Villain}), as the expectation value of the $z$-component of
each spin is now substantially different from zero. On the other hand, the
Holstein-Primakoff transformation (HPT)~\cite{HolsteinP40}
\begin{eqnarray}
&\,&\sqrt{2}\hat{S}^+=\sqrt{S+\widetilde{S}-\hat{z}^2}~(\hat{q}+i\hat{p})
~,\nonumber\\
&\,&\hat{S}^-=(\hat{S}^+)^\dagger
~,~~~\hat{S}^z=\widetilde{S}-\hat{z}^2~,
\label{e.HolsteinP}
\end{eqnarray}
and the Dyson-Maleev one (DMT)~\cite{DysonM56}
\begin{eqnarray}
&\,&\sqrt{2}\hat{S}^+=\hat{q}+i\hat{p}
~,~~~\hat{S}^-=(\hat{q}-i\hat{p})(S+\widetilde{S}-\hat{z}^2)~,\nonumber \\
&\,&\hat{S}^z=\widetilde{S}-\hat{z}^2~,
\label{e.DysonM}
\end{eqnarray}
with $\hat{z}^2\equiv (\hat{q}^2+\hat{p}^2)/2$, do both suggest the
$z$-component to be the privileged one for alignment, thus fitting to the
$|\mu|<1$ case. 

The isotropic, or nearly isotropic, case is the most difficult to be
treated, as no part of the spherical phase-space may be generically chosen
to be described better than another; nevertheless, when the ground state
is at least characterized by long-range order with an alignment axis, 
the HP and DM transformations are both seen to work well in the
PQSCHA scheme. The reason is that the boson picture only serves to derive
the pure-quantum renormalizations, while the symmetry is thereafter
restored in ${\cal H}_{\rm eff}$ so that it is sufficient to assume a
local alignment axis.

\section{The easy-plane case}
\label{s.easypl}

The Hamiltonian Eq.~(\ref{e.H}) in the easy-plane case takes the form
\begin{equation}
\hat{\cal H}=-J\sum_{<{\bf ij}>}
\left[\left(
\hat{S}^x_{\bf i}\hat{S}^x_{\bf j}+\hat{S}^y_{\bf i}\hat{S}^y_{\bf j}
\right)
+\lambda \hat{S}^z_{\bf i}\hat{S}^z_{\bf j}\right]
\label{e.easyplH}
\end{equation}
with $\lambda\in (-1,1)$. The antiferromagnetic sector is mapped in the 
$\lambda\in (-1,0]$ region of Eq.~(\ref{e.easyplH}) by the canonical
transformation $(\hat{S}^x_{\bf i},\hat{S}^y_{\bf i},\hat{S}^z_{\bf i})
\rightarrow
\left( (-1)^{\bf i}\hat{S}^x_{\bf i},
       (-1)^{\bf i}\hat{S}^y_{\bf i},
       \hat{S}^z_{\bf i}
\right)$ where $(-1)^{\bf i}\equiv (-1)^{i_1+i_2}$; ferro-
and antiferromagnets may be thus simultaneously studied.

The effective Hamiltonian relative to Eq.~(\ref{e.easyplH}) is
\begin{equation}
{\cal H}_{\rm eff}=-J\widetilde{S}^2~j_{\rm eff}\sum_{<{\bf ij}>}
\left[\left(s^x_{\bf i}s^x_{\bf j}+s^y_{\bf i}s^y_{\bf j}\right)+
\lambda_{\rm eff}s^z_{\bf i}s^z_{\bf j}\right]+{\cal G}~,
\label{e.easyplHeff}
\end{equation}
where the renormalization parameters $j_{\rm eff}(t,\lambda,S)$,\break
$\lambda_{\rm eff}(t,\lambda,S)$, and ${\cal G}(t,\lambda,S)$ have
different expressions depending on the $\lambda$-region considered. 

In the $\lambda\ll 1$ case, the VT
Eqs.~(\ref{e.Villain}) may be safely used, and give~\cite{CTVVprb95}
\begin{eqnarray}
j_{\rm eff}^{\rm V}&=&(1-{1\over 2}D_\perp)^2~e^{-{\cal D}_\parallel/2}~,
\\
\lambda_{\rm eff}^{\rm V}&=&\lambda
(1-{1\over 2}D_\perp)^{-1}~e^{{\cal D}_\parallel/2}~,\\
{\cal G}^{\rm V}&=&J\widetilde{S}^2N
\left[t\Gamma - {t\over 2}\ln\Big(1-{1\over 2}D_\perp\Big)
      -\Theta^{\rm V}\right]~,
\label{e.epGV}
\end{eqnarray}
where the renormalization coefficients $D_\perp$ and ${\cal D}_\parallel$
represent the effects of in-plane and out-of-plane pure-quantum
fluctuations, respectively, and have the form
\begin{eqnarray}
D_\perp&=&{1\over 2\widetilde{S}N}\sum_{\bf k}
          {b_{\bf k}\over a_{\bf k}}{\cal L}_{\bf k}~,\\
{\cal D}_\parallel&=&{1\over 2\widetilde{S}N}\sum_{\bf k}
          (1-\gamma_{\bf k}){a_{\bf k}\over b_{\bf k}}{\cal L}_{\bf k}~,
\end{eqnarray}
with 
\begin{eqnarray}
a^2_{\bf k}&=&4e^{-{\cal D}_\parallel/2}(1-\lambda_{\rm eff}\gamma_{\bf k})~,
\\
b^2_{\bf k}&=&4(1-{1\over 2}D_\perp)^2
              e^{-{\cal D}_\parallel/2}(1-\gamma_{\bf k})~,
\end{eqnarray}
${\cal L}_{\bf k}=(\coth f_{\bf k}-1/f_{\bf k})$, 
$\Gamma=N^{-1}\sum_{\bf k}\ln(\sinh f_{\bf k}/f_{\bf k})$,
$f_{\bf k}=a_{\bf k}b_{\bf k}/(2\widetilde{S} t)$,
$\gamma_{\bf k}=\sum_{\bf d}\cos({\bf k}\cdot{\bf d})/4$, 
${\bf d}=(\pm 1,\pm 1)$, and $\bf k$ is a wave vector in the first
Brillouin zone. The uniform coefficient $\Theta^{\rm V}$ is 
$$
\Theta^{\rm V}=e^{-{\cal D}_\parallel/2}
\left(2D_\perp+(1-D_\perp){\cal D}_\parallel\right)~.
$$

Close to the isotropic limit, $\lambda\lesssim 1$, 
the HPT must be used: because of the residual easy-plane
character of the model, the privileged axis, i.e. the
quantization one, must be chosen in the easy plane; in particular, we
choose the $y$ direction as the local alignment one.
A quite complicated PQSCHA implementation finally leads to the
following expressions~\cite{CCTVV98}:
\begin{eqnarray}
j_{\rm eff}^{\rm HP}&=&\theta^4~,\\
\lambda_{\rm eff}^{\rm HP}&=&\lambda+{d\over\theta^2}(1-\lambda)+
                             {D^{pp}\over\theta^2}(1-\lambda)^2~, \\
{\cal G}^{\rm HP}&=&J\widetilde{S}^2N\left[t\Gamma +
{\theta^2\over N\widetilde{S}}\sum_{\bf k}a_{\bf k}b_{\bf k}{\cal L}_{\bf k}
-\Theta^{\rm HP}\right]~,
\label{e.epGHP}
\end{eqnarray}
where the temperature and spin dependent renormalization coefficients
are defined as follows:
\begin{eqnarray}
&\,&\theta^2=1-{1\over 2}{\cal D}+{1\over 2}d~~,~~
d={1\over 2}(D^{qq}-\lambda_{\rm eff}D^{pp})~,\\
&\,&{\cal D}={1\over N\widetilde{S}}\sum_{\bf k}
           (1-\gamma_{\bf k}){a_{\bf k}\over b_{\rm k}}{\cal L}_{\bf k}~,
\end{eqnarray}
and 
$$
D^{pp}={1\over 2\widetilde{S}N}\sum_{\bf k}
         {a_{\bf k}\over b_{\rm k}}\gamma_{\bf k}{\cal L}_{\bf k}~~,~~
D^{qq}={1\over 2\widetilde{S}N}\sum_{\bf k}
         {b_{\bf k}\over a_{\rm k}}\gamma_{\bf k}{\cal L}_{\bf k}~,
$$
with dimensionless frequencies  
\begin{eqnarray}
a^2_{\bf k}&=&4\theta^2(1-\lambda_{\rm eff}\gamma_{\bf k})~,\\
b^2_{\bf k}&=&4\theta^2(1-\gamma_{\bf k})~.
\end{eqnarray}
The uniform coefficient $\Theta^{\rm HP}$ is 
\begin{eqnarray*}
\Theta^{\rm HP}&=&
D^{qq}(1-3D^q-D^p)+\lambda D^{pp}(3-3D^p-D^q)+\\
&+&2(D^{qq})^2+2(D^{pp})^2-(D^q+D^p)^2+2(D^q+D^p)~,
\end{eqnarray*}
with
$$
D^p={1\over 2\widetilde{S}N}\sum_{\bf k}
         {a_{\bf k}\over b_{\rm k}}{\cal L}_{\bf k}~~,~~ 
D^q={1\over 2\widetilde{S}N}\sum_{\bf k}
         {b_{\bf k}\over a_{\rm k}}{\cal L}_{\bf k}~.
$$

The renormalization of the exchange energy by the factor $j_{\rm eff}<1$,
and the weakening of the easy-plane anisotropy
($|\lambda|\leq|\lambda_{\rm eff}|$), are seen to be, both for strong and
weak anisotropy, the result of the cooperative effect of in-plane and
out-of plane pure-quantum fluctuations. 
It is worthwhile noticing that, in
the isotropic limit $|\lambda|=1$, the effective Hamiltonian 
(\ref{e.easyplHeff}), determined by the HPT,
coincides
with the one found directly for the isotropic model by the DMT.

\begin{figure}
\begin{center}
\includegraphics[bbllx=25mm,bblly=30mm,bburx=200mm,bbury=250mm,%
width=70mm,angle=90]{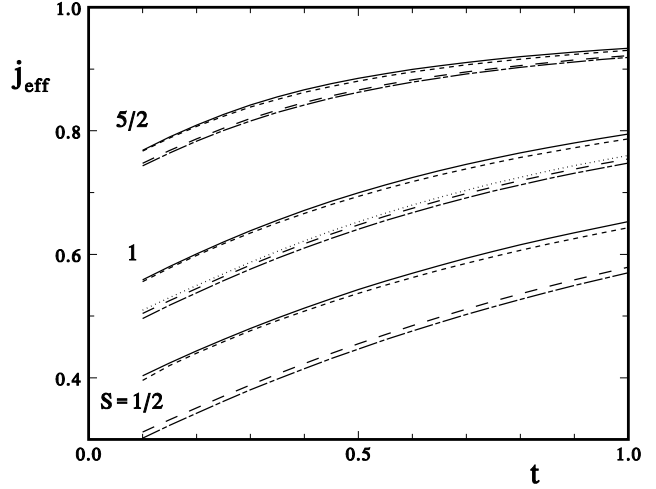}
\end{center}
\caption{Renormalization parameter $j_{\rm eff}$ {\it vs} $t$ for 
the easy-plane model with $\lambda=0$ (full curve), $0.5$ (dashed), $0.7$ 
(long-dashed), $0.9$ (dash-dotted), and $S=1/2$, $1$, $5/2$.
The dotted line is $j_{\rm eff}$ for $S=1$, $\lambda=0.5$, as obtained by 
the HPT.}
\label{f.epj}
\end{figure}

\begin{figure}
\begin{center}
\includegraphics[bbllx=25mm,bblly=30mm,bburx=200mm,bbury=250mm,%
width=70mm,angle=90]{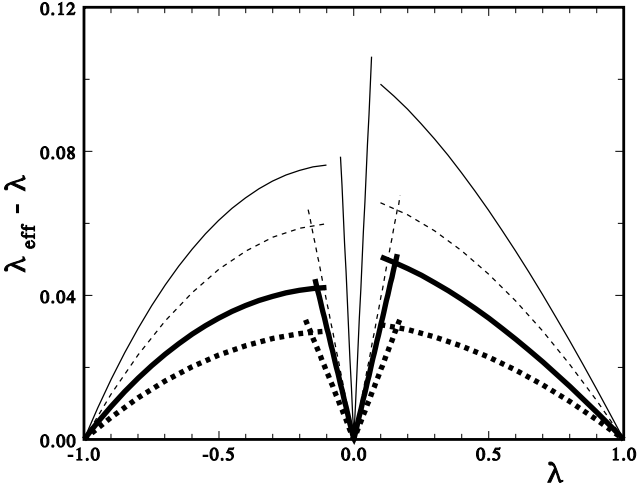}
\end{center}
\caption{Difference $(\lambda_{\rm eff}-\lambda)$ {\it vs}
$\lambda$ for the easy-plane model with $S=1/2$ (thin) and $1$ (thick), at
$t=0.5$ (full) and $1$ (dashed). For each $t$ and $S$, both the 
HP (curves passing through $(\pm 1,0)$) and the V ( curves passing
through the origin) results are shown.}
\label{f.eplam}
\end{figure}

In Fig.~\ref{f.epj} we show the renormalization
parameters $j_{\rm eff}$ vs. $t$, for different values of $S$ and
$\lambda$; the two most (least) anisotropic cases are treated by the VT
(HPT). For $S=1$ and $\lambda=0.5$, $j_{\rm eff}$ as obtained by the HPT
is also shown (dotted line): the difference with the VT result (dashed
line) is remarkable, testifying that both transformations cannot give very
accurate results in such intermediate region. However, as we will see in
Section \ref{s.global}, this does not dramatically affect the
critical temperature estimation.

The difference $(\lambda_{\rm eff}-\lambda)$ is shown in
Fig.~\ref{f.eplam} as a function of $\lambda$, for different values of $S$
and $t$. 
For each couple $(t,S)$ the curves passing through the origin are obtained
by the VT, while those passing through the point $(-1,0)$ or $(1,0)$ are
obtained by the HPT in the antiferro- or ferromagnetic case, respectively;
the two types of results do not smoothly connect to each other, but
nevertheless cover the whole $[-1,1]$ interval and show a behaviour of the
same type of that found in the easy-axis case (see Fig.\ref{f.eamu}),
where the problem of using two different spin boson transformations is not
present. As expected, the HP curves display the correct $|\lambda|\to 1$
behaviour (i.e. $|\lambda_{\rm eff}|^{\rm HP}\to 1$), while 
$\lambda_{\rm eff}^{\rm V}$ is seen to be proportional to
$\lambda$, when $\lambda\to 0$.

As mentioned in the Introduction, the easy-plane models are
characterized by a critical phase, with vanishing magnetization and
diverging correlation length, extending in the
temperature region $0<t<t_{\rm BKT}$; thermodynamic quantities that may be
significantly studied in such phase are, for instance, the free energy
and the specific heat. To this respect, notice that the uniform term
${\cal G}$, Eqs.~(\ref{e.epGV}) and (\ref{e.epGHP}), not entering
the statistical averages' calculation, is
essential in the evaluation of the partition function ${\cal Z}$.

In the paramagnetic phase $t>t_{\rm BKT}$, one of the
fundamental thermodynamic quantities to be investigated, is the in-plane
correlation length defined by the asymptotic behaviour of
the correlation functions
\begin{equation}
G(r)\equiv{\eta_{\bf r}\over\widetilde{S}^2}
\langle
\hat{S}^x_{\bf i}\hat{S}^x_{{\bf i}+{\bf r}}+
\hat{S}^y_{\bf i}\hat{S}^y_{{\bf i}+{\bf r}}
\rangle
\mathop{\sim}\limits_{r\to\infty}
 e^{-{r\over \xi}}~
\end{equation}
with ${\bf r}=(r_1,r_2)$ any vector on the square lattice, 
$r\equiv |{\bf r}|$ and $\eta_{\bf r}=1$ or
$(-1)^{r_1+r_2}$ in the ferro- or antiferromagnetic case.

By using Eq.~(\ref{e.aveO}) the following PQSCHA expression is found
\begin{equation}
G(r)=
\Delta_{\bf r}\langle s^x_{\bf i}s^x_{{\bf i}+{\bf r}}+
                 s^y_{\bf i}s^y_{{\bf i}+{\bf r}}\rangle_{\rm eff}~,
\label{e.easyplGr}
\end{equation}
where 
$\langle~\cdot~\rangle_{\rm eff}\equiv 
{\cal Z}^{-1}\int\,d^{\scriptscriptstyle N}{\bf s}(~\cdot~)
e^{-\beta{\cal H}_{\rm eff}}$ 
is the classical-like statistical average defined by the effective
Hamiltonian. 

When the VT is used, i.e. in the strongly easy-plane case, 
Eq.~(\ref{e.easyplGr}) holds for all values of $r$, with
\begin{equation}
\Delta_{\bf r}^{\rm VT}=(1-{1\over 2}{\cal D}_\perp)^2 
e^{-{\cal D}_\parallel/2} e^{-D_{\bf r}}~,
\label{e.epDeltaV}
\end{equation}
and
$$
D_{\bf r}={1\over \widetilde{S}N}\sum_{\bf k}
{a_{\bf k}\over b_{\bf k}} \cos ({\bf k}\cdot {\bf r}) {\cal L}_{\bf k}~.
$$

In the nearly isotropic region, the use of the HPT leads to a much more
complicated general expression for $G(r)$, that nonetheless goes to 
expression (\ref{e.easyplGr}) with
coefficient
\begin{equation}
\Delta_{\bf r}^{\rm HP}\to\left(1-{1\over 2}(D^q+D^p)\right)
                        \left(1-{1\over 8}(5D^q+3D^p)\right)~,
\label{e.epDeltaHP}
\end{equation}
for large $r$~.
From Eqs.~(\ref{e.epDeltaV}) and (\ref{e.epDeltaHP}), 
the renormalization coefficient $\Delta_{\bf r}$ is seen to converge to a
finite value when $r\to\infty$;
this property, as we will see in Section \ref{s.global},
is of fundamental importance when the critical behaviour of the model is
under investigation. In particular it means that the correlation length of
the quantum model is related with that of its classical counterpart by the
relation
\begin{equation}
\xi(t,\lambda,S)=\xi^{\rm cl}(t_{\rm eff},\lambda_{\rm eff})
\label{e.easyplxi}
\end{equation}
with $t_{\rm eff}\equiv t/j_{\rm eff}$. This relation allows us to
evaluate the correlation length of a quantum model with spin $S$, at a
given temperature $t$, by performing a classical MC
simulation on a model with anisotropy $\lambda_{\rm eff}$, at
the effective temperature $t_{\rm eff}$;
a particularly simple case is that of the $XY$ model
$\lambda=\lambda_{\rm eff}=0$, whose correlation length may be directly
determined in terms of the classical one, by a simple temperature scaling.  
Fig.~\ref{f.epxi} shows the correlation length versus $t$, for $S=1$
and $\lambda=0$, $-0.5$, $-0.999$, and $-1$. The classical MC data for
$\lambda=0$ and $\lambda=0.999$, Refs.~\cite{CTV95} and \cite{CCTVV98}
respectively, are also shown.  

\begin{figure}
\begin{center}
\includegraphics[bbllx=25mm,bblly=30mm,bburx=200mm,bbury=250mm,%
width=70mm,angle=90]{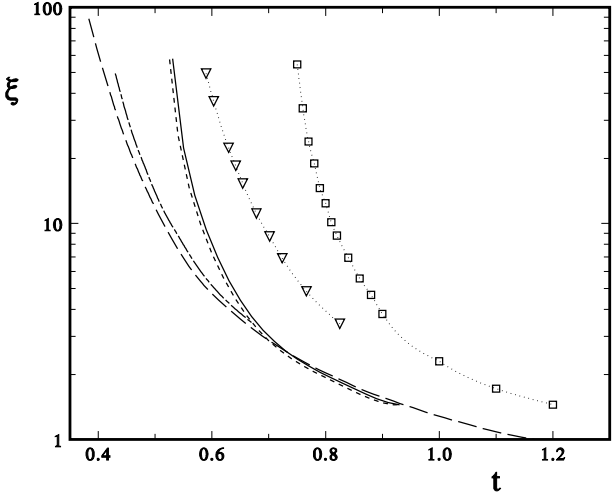}
\end{center}
\caption{Correlation length $\xi$ {\it vs} $t$ for 
the easy-plane model with $S=1$ and $\lambda=0$ (full), $-0.5$ (dashed),
$-0.999$ (dash-dotted), and $-1$ (long-dashed). Symbols are 
classical MC data for $\lambda=0$ (squares~\cite{CTV95}) and
$\lambda=0.999$ (triangles~\cite{CCTVV98}).} 
\label{f.epxi}
\end{figure}

\section{The easy-axis case}
\label{s.easyax}

The Hamiltonian Eq.~(\ref{e.H}) in the easy-axis case takes the form
\begin{equation}
\hat{\cal H}=-J\sum_{<{\bf ij}>}
\left[\mu\left(
\lambda\hat{S}^x_{\bf i}\hat{S}^x_{\bf j}+\hat{S}^y_{\bf i}\hat{S}^y_{\bf
j}
\right)
+\hat{S}^z_{\bf i}\hat{S}^z_{\bf j}\right]
\label{e.easyaxH}
\end{equation}
with $\mu\in (-1,1)$ and $\lambda={\rm sign}(\mu)$. 
The antiferromagnetic sector is mapped in the 
$\mu\in (-1,0]$ region of Eq.~(\ref{e.easyaxH}) by the canonical
transformation \break
$(\hat{S}^x_{\bf i},\hat{S}^y_{\bf i},\hat{S}^z_{\bf i})\rightarrow\left( (-1)^{\bf i}\hat{S}^x_{\bf i},\hat{S}^y_{\bf i},(-1)^{\bf i}\hat{S}^z_{\bf i}\right)$.
Because of the easy-axis character of the model, the HP and DM
transformations may be safely used, and actually give the same
results; the latter, however, may be preferred because of its simpler
structure. 

The PQSCHA effective Hamiltonian relative to Eq.(\ref{e.easyaxH}) is
found to be~\cite{CRTVV00}
\begin{equation}
{\cal H}_{\rm eff}=-J\widetilde{S}^2~j_{\rm eff}\sum_{<{\bf ij}>}
\left[\mu_{\rm eff}\left(
\lambda s^x_{\bf i}s^x_{\bf j}+s^y_{\bf i}s^y_{\bf j}\right)
+s^z_{\bf i}s^z_{\bf j}\right]+{\cal G}~,
\label{e.easyaxHeff}
\end{equation}
where $j_{\rm eff}(t,\mu,S)$, $\mu_{\rm eff}(t,\mu,S)$, and
${\cal G}(t,\mu,S)$ are
\begin{eqnarray}
&\,&j_{\rm eff}=\theta^4_\parallel~~,~~
\mu_{\rm eff}=\mu{\theta^2_\perp\over\theta^2_\parallel}~,\\
&\,&{\cal G}=
2J\widetilde{S}^2(1-\theta^4_\parallel)+J\widetilde{S}^2\Gamma~,
\end{eqnarray}
with
\begin{eqnarray}
&\,&\theta^2_\parallel=1-{{\cal D}_\parallel\over 2}~~,~~
\theta^2_{\perp}=1-{{\cal D}_\perp\over 2}~,
\label{e.theta}\\
&\,&\Gamma=t\sum_{\bf k}
\ln\left({\sinh f_{\bf k}\over \theta^2_\perp f_{\bf k}}\right)~,
\label{e.Gamma}
\end{eqnarray}
and the coefficients
\begin{eqnarray}
{\cal D}_\parallel&=&{1\over N\widetilde{S}}\sum_{\bf k}
{a_{\bf k}\over b_{\bf k}}\left(1-\mu\lambda\gamma_{\bf k}\right)
{\cal L}_{\bf k}~,\nonumber\\
{\cal D}_\perp&=&{1\over N\widetilde{S}}\sum_{\bf k}
{a_{\bf k}\over b_{\bf k}}\left(1-{\gamma_{\bf k}\over\mu\lambda}\right)
{\cal L}_{\bf k}
\label{e.D}
\end{eqnarray}
are self-consistently determined by solving Eqs.~(\ref{e.theta})
and (\ref{e.D}), with dimensionless frequencies
\begin{eqnarray}
a^2_{\bf k}&=&
4\left(\theta^2_\parallel-\mu\theta^2_\perp\gamma_{\bf k}\right)~,\\
b^2_{\bf k}&=&
4\left(\theta^2_\parallel-\mu\lambda\theta^2_\perp\gamma_{\bf k}\right)~.
\label{e.akbk}
\end{eqnarray}
As in the easy-plane case, quantum renormalizations wea\-ken both the
energy scale ($j_{\rm eff}<1$) and the anisotropy ($|\mu_{\rm eff}|>\mu$).
In Fig.~\ref{f.eaj} we show the renormalization
parameter $j_{\rm eff}$ vs. $t$ for different values of $S$ and
$\mu$. In Fig.~\ref{f.eamu}, where the difference $\mu_{\rm
eff}-\mu$ is shown as a function of $\mu$, we see that, at variance
with the easy-plane case (see Fig.~\ref{f.eplam}), the almost
isotropic regime is smoothly connected with the strongly axial
one, testifying to the suitableness of the DMT in treating the
whole easy-axis subclass of models.

\begin{figure}
\begin{center}
\includegraphics[bbllx=25mm,bblly=30mm,bburx=200mm,bbury=250mm,%
width=70mm,angle=90]{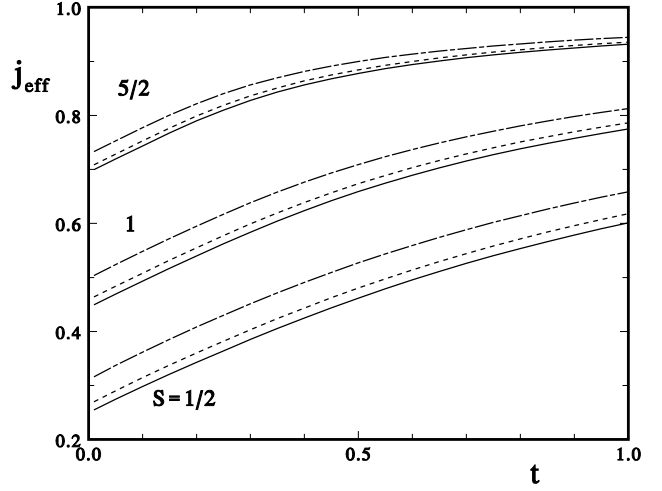}
\end{center}
\caption{Renormalization parameter $j_{\rm eff}$ {\it vs} $t$ for 
the easy-axis model with $\mu=0$ (full), $0.5$ (dashed), $0.9$
(dashed), and $S=1/2$, $1$, $5/2$.}
\label{f.eaj}
\end{figure}

\begin{figure}
\begin{center}
\includegraphics[bbllx=25mm,bblly=30mm,bburx=200mm,bbury=250mm,%
width=70mm,angle=90]{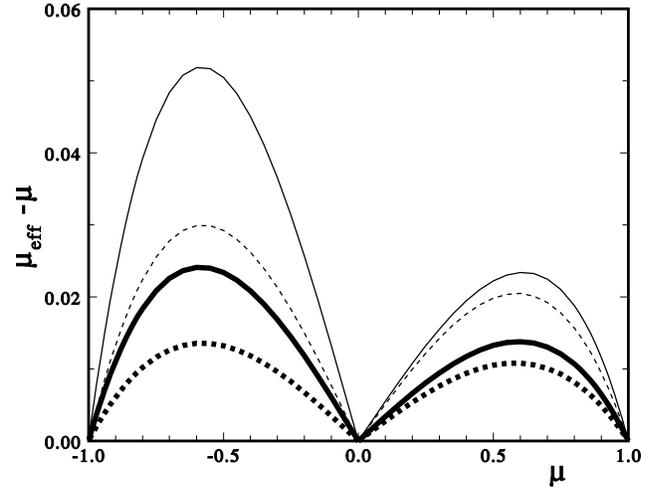}
\end{center}
\caption{Difference $(\mu_{\rm eff}-\mu)$ {\it vs} $\mu$ for 
the easy-axis model with $S=1/2$ (thin) and  $1$ (thick), at
$t=0.5$ (full) and $1$ (dashed).}
\label{f.eamu}
\end{figure}

When $\mu$ vanishes the renormalization coefficient ${\cal D}_\perp$ 
may be shown to keep finite, so that $\mu_{\rm eff}\to 0$, and the $Z$
model, with a single optical mode 
$a^2_{\bf k}=b^2_{\bf k}=4\theta^2_\parallel$,
$\forall {\bf k}$, is recovered; for $S=1/2$ such model coincides
with the Ising one.

The easy-axis case is
characterized by a finite temperature phase-transition towards a fully
ordered phase; many thermodynamic properties may be hence studied both
above and below the critical temperature $t_{\rm I}$, and,
in particular, the PQSCHA expression for the magnetization along the easy axis 
$m\equiv\sum_{\bf i}\eta_{\bf i}\langle\hat{S}^z_{\bf i}\rangle
/N\widetilde{S}$ is
\begin{equation}
m=\theta^2_\parallel\langle s^z_{\bf i}\rangle_{\bf eff}~.
\label{e.easyaxm}
\end{equation}

The most significant quantity in the paramagnetic phase $t>t_{\rm I}$ is
the correlation length, here defined by the asymptotic behaviour of the
isotropic correlation functions
\begin{equation}
G(r)\equiv{\eta_{\bf r}\over\widetilde{S}^2}
\langle
\hat{\bf S}_{\bf i}\cdot\hat{\bf S}_{{\bf i}+{\bf r}}\rangle
\mathop{\sim}\limits_{r\to\infty}
 e^{-{r\over\xi}}~;
\end{equation}
the PQSCHA expression for $G(r)$ is
\begin{equation}
G(r)=\theta^4_{\bf r}
\langle{\bf s}_{\bf i}\cdot{\bf s}_{{\bf i}+{\bf r}}\rangle_{\rm eff}~,
\label{e.Gr}
\end{equation}
where $\theta^2_{\bf r}=1-{\cal{D}}_{\bf r}/2$ and
$$
{\cal{D}}_{\bf r}={1\over N\widetilde{S}}\sum_{\bf k}{a_{\bf k}\over b_{\bf k}}
\left(1-\cos({\bf k}{\cdot}{\bf r})\right){\cal L}_{\bf k}
$$
is a further site-dependent renormalization coefficient.
For increasing $r$, as in the easy-plane case,  the coefficient
$\theta^4_{\bf r}$ keeps finite and rapidly converges to a uniform term:
the asymptotic behaviour of $G(r)$ is hence determined by that of the
classical-like correlation functions 
$\langle{\bf s}_{\bf i}\cdot{\bf s}_{{\bf i}+{\bf r}}\rangle_{\rm eff}$,
and the relation 
\begin{equation}
\xi(t,\mu,S)=\xi^{\rm cl}(t_{\rm eff},\mu_{\rm eff})
\label{e.easyaxxi}
\end{equation}
may be used. 
As in the easy-plane case, results for the quantum
reference ($Z$) model are obtained by a simple temperature scaling of
the classical data; we have hence performed MC simulations on the
classical $Z$ model, and produced results for the quantum
model with $S=1/2$, $1$ and $5/2$. The corresponding magnetization and
correlation-length curves are shown in Figs.\ref{f.Zm} and \ref{f.Zxi},
respectively. As for the latter, notice the much sharper divergence with
respect to the $XY$ model (compare the dashed curve in Fig.\ref{f.Zxi}
with the full curve of Fig.\ref{f.epxi}). 

\begin{figure}
\begin{center}
\includegraphics[bbllx=25mm,bblly=30mm,bburx=200mm,bbury=250mm,%
width=70mm,angle=90]{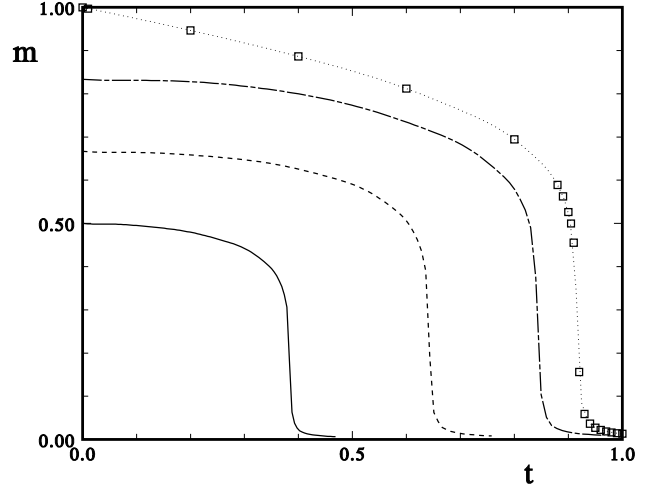}
\end{center}
\caption{Magnetization curves for the $Z$ model
with $S=1/2$ (full), $1$ (dashed), $5/2$ (dash-dotted), and
$\infty$ (dotted); symbols are our classical MC data.}
\label{f.Zm}
\end{figure}

\begin{figure}
\begin{center}
\includegraphics[bbllx=25mm,bblly=30mm,bburx=200mm,bbury=250mm,%
width=70mm,angle=90]{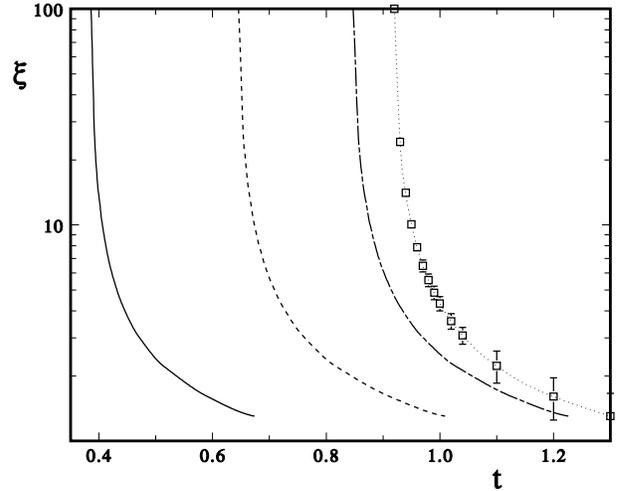}
\end{center}
\caption{Correlation length for the $Z$ model with
$S=1/2$, $1$, $5/2$, and $\infty$ (curves and symbols as in previous
figure).}
\label{f.Zxi}
\end{figure}

\section{Transition temperatures}
\label{s.global}

The PQSCHA expressions for the effective Hamiltonian and the
correlation functions contain two essential pieces of information:
({\it a}) the effective Hamiltonian keeps the symmetry properties of the
original quantum model; ({\it b}) the quantum correlation functions 
share the asymptotic behaviour with their classical counterpart.
These two statements not only lead us to the relations
 (\ref{e.easyplxi}) and (\ref{e.easyaxxi}), but also allow us
to assert that the critical behaviour of the quantum model is that of
its effective classical counterpart, and that, for the critical
temperature, the relation
\begin{equation} 
t_c(\nu,S)=j_{\rm eff}(t_c,\nu,S)~t^{\rm cl}_c(\nu_{\rm eff}(t_c,\nu,S))~
\label{e.tc}
\end{equation}
holds.

Due to the temperature dependence of $j_{\rm eff}$ and $\nu_{\rm eff}$,
Eq.~(\ref{e.tc}) is quite involute; it may be nonetheless numerically
solved if an analytical expression for $t_c^{\rm cl}(\nu)$ is available.
Such expression does not actually exist, neither in the easy-plane nor in
the easy-axis case; we have hence collected a set of classical MC
data properly distributed in the $[0,1)$ interval, in order to 
determine a precise interpolating function $\tau_c^{\rm cl}(\nu)$.

In the easy-plane case, classical MC data have been taken from
Ref.~\cite{CTV95} and interpolated by the function
\begin{eqnarray}
&\,&\tau_{\rm BKT}^{\rm cl}(\lambda)=
a\left(1+b\lambda^2 + c\lambda^4 - d\ln(1-\lambda^2)\right)^{-1}~,
\label{e.tBKT}\\
&\,&a=0.695~,~~b=-0.012~,~~c=0.032~,~~d=0.0648~,\nonumber
\end{eqnarray}
where the $\lambda\to -\lambda$ invariance has been imposed.

In the easy-axis case, sources of the classical MC data have been
Refs.~\cite{SerenaGL93} and \cite{GouveaEtal99}, as well as our new data
$t_{\rm I}^{\rm cl}(0.5)=0.88\pm 0.01$ and $t_{\rm I}^{\rm
cl}(0.8)=0.81\pm 0.01$; the resulting interpolating function is
\begin{eqnarray}
&\,&\tau_{\rm I}^{\rm cl}(\mu)=
a\left(1+b\mu^2 + c\mu^4 - d\ln(1-\mu^2)\right)^{-1}~,\label{e.tI}\\
&\,&a=0.917~,~~b=0.068~,~~c=0.097~,~~d=0.0636~,\nonumber
\end{eqnarray}
where, again, the constraint 
$t_{\rm I}^{\rm cl}(\mu)=t_{\rm I}^{\rm cl}(-\mu)$ has been imposed.

Equations (\ref{e.tBKT}) and (\ref{e.tI}) should only be considered as
reasonable interpolating functions of the available data; in particular, 
they do not reproduce the correct asymptotic behaviour in the $\nu\to 1$
isotropic limit.
Such asymptotic behaviours, on the other hand, are not accessible by
numerical techniques and are actually determined by Renormalization-Group
approaches.
However, all of the numerical and experimental data 
 are in the $\nu$-region where $t_c(\nu)$ is sensibly different from
zero, as seen in Figs.~\ref{f.tBKT} and \ref{f.tI}, i.e. where the
above mentioned asymptotic behaviours have not set in yet.

Once $t_c^{\rm cl}(\nu)\simeq\tau_c^{\rm cl}(\nu)$ has been determined, 
Eq.~(\ref{e.tc}) may be numerically solved thus giving the
value of the quantum critical temperature, for any value of $\nu$ and $S$,
in a few seconds on a standard PC. 

In Fig.~\ref{f.tBKT} we show the
resulting curves for $S=1/2$, $1$, $5/2$, and $\infty$ in the easy-plane
case: for each spin, the whole $(-1,1)$ interval is covered by two
different types of curves, corresponding to the use of 
$j^{\rm V}_{\rm eff}$,$\lambda^{\rm V}_{\rm eff}$ and
$j^{\rm HP}_{\rm eff}$,$\lambda^{\rm HP}_{\rm eff}$ in the strongly
easy-plane and nearly isotropic regions, respectively.
The available quantum  MC data~\cite{DingM90,Ding92} are also shown, and
agree very well with
our results, despite
being in the region where, because of the strong quantum fluctuations,
the use of the PQSCHA becomes more delicate; unfortunately no
experimental data exist, to our knowledge, for $t_{\rm BKT}$.

In Fig.~\ref{f.tI} the transition temperature $t_{\rm I}(\mu,S)$ is
shown for
$S=1/2$, $1$, $5/2$, and $\infty$; in this case, there exist several
experimental data, for the real compounds 
K$_2$NiF$_4$ ($\mu=-0.996$, $S=1$, $t_{\rm I}^{\rm exp}=0.416$)
~\cite{deWijnWW73,SkalyoEtal69},
Rb$_2$NiF$_4$ ($\mu=-0.982$, $S=1$, $t_{\rm I}^{\rm exp}=0.477$) 
~\cite{NagataT74},
K$_2$MnF$_4$ ($\mu=-0.995$, $S=5/2$, $t_{\rm I}^{\rm exp}=0.553$)
~\cite{deWijnWW73,BirgeneauGS73},
Rb$_2$MnCl$_4$ ($\mu=-0.997$, $S=5/2$, $t_{\rm I}^{\rm exp}=0.558$)
~\cite{SchroederEtal80}, and
Rb$_2$MnF$_4$ ($\mu=-0.994$, $S=5/2$, $t_{\rm I}^{\rm exp}=0.575$)
~\cite{deWijnWW73,CowleyEtal77},
whose agreement with our results is very good, despite being in 
the most difficult region to be studied, i.e. in the nearly isotropic one.

\begin{figure}
\begin{center}
\includegraphics[bbllx=25mm,bblly=30mm,bburx=200mm,bbury=250mm,%
width=70mm,angle=90]{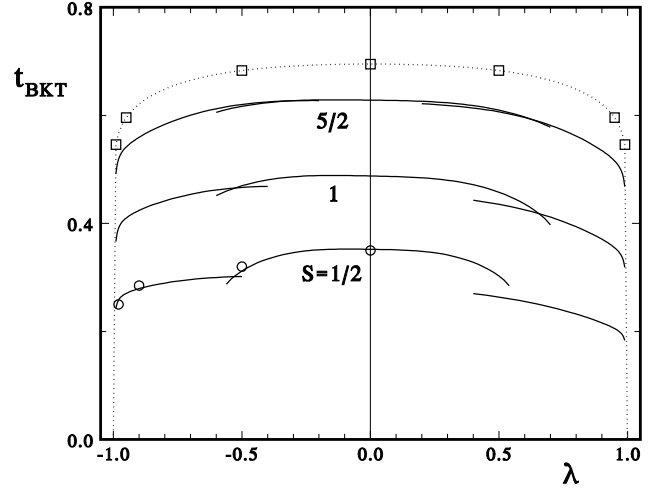}
\end{center}
\caption{BKT critical temperature {\it vs} $\lambda$ for the easy-plane
model with $S=1/2$, $1$, $5/2$. 
Symbols are the quantum MC data for $S=1/2$
(circles~\protect\cite{DingM90,Ding92}), 
and the classical MC data (squares~\protect\cite{CTV95}) used to construct 
$\tau_{\rm BKT}^{\rm cl}(\lambda)$, shown as dotted line.}
\label{f.tBKT}
\end{figure}

\begin{figure}
\begin{center}
\includegraphics[bbllx=25mm,bblly=30mm,bburx=200mm,bbury=250mm,%
width=70mm,angle=90]{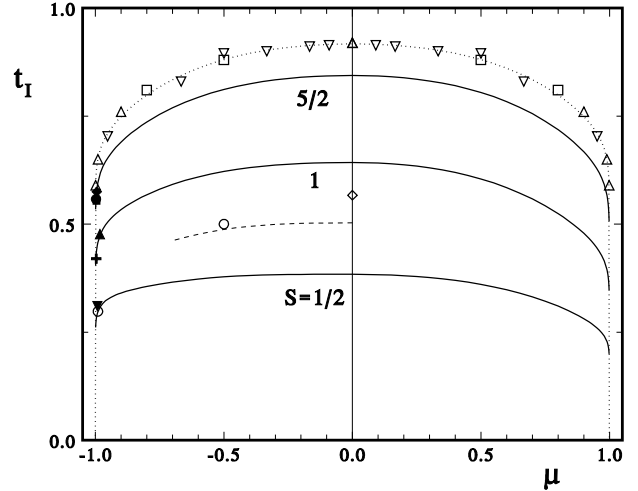}
\end{center}
\caption{Ising critical temperature {\it vs} $\mu$ for the easy-axis model
with $S=1/2$, $1$, $5/2$. Symbols are
quantum MC data~\protect\cite{Ding90} for $S=1/2$ (open circles),
experimental
data (filled symbols) for several real compounds  (see Fig.\ref{f.tIing}
for details), and the classical MC data (upwards triangles
~\protect\cite{GouveaEtal99}, 
downwards triangles~\protect\cite{SerenaGL93}, circles from our new
simulations) 
used to construct $\tau_{\rm I}^{\rm cl}(\mu)$, shown as dotted line; the
open diamond is the exact result for the Ising model; see text for the
dashed line.}
\label{f.tI}
\end{figure}

\begin{figure}
\begin{center}
\includegraphics[bbllx=25mm,bblly=30mm,bburx=200mm,bbury=250mm,%
width=70mm,angle=90]{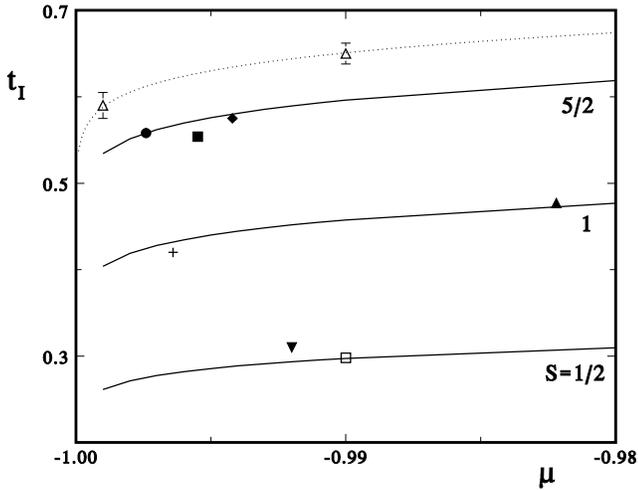}
\end{center}
\caption{Ising critical temperature for the antiferromagnetic easy-axis
model, near the isotropic limit. Curves and open symbols are as in
Fig.\ref{f.tI};
Filled symbols are experimental data for the compounds
YBa$_2$Cu$_3$O$_6.1$ (downwards triangle), K$_2$NiF$_4$ (cross), Rb$_2$Ni
F$_4$ (upwards triangle), Rb$_2$MnCl$_4$ (circle), Rb$_2$MnF$_4$
(diamond); see text for references.} 
\label{f.tIing} 
\end{figure}

The $S=1/2$ easy-axis antiferromagnetic case deserves a few more comments,
as a quite well distributed scan of the $\mu\in[-1,0)$ region may be
performed by putting together the exact result for the Ising
model~\cite{Onsager44}, 
$t_{\rm I}(0,1/2)=0.567$, the quantum MC data by Ding~\cite{Ding90}, 
$t_{\rm I}(0.5,1/2)=0.5$,
$t_{\rm I}(0.99,1/2)=0.298$, and the experimental data for the real
compound YBa$_2$Cu$_3$O$_{6.1}$ 
($\mu\simeq 0.992$, $S=1/2$, $t_{\rm I}^{\rm exp}=0.310$)
~\cite{RossatMignodEtal91}. 
As for the PQSCHA, we know that its results are
quantitatively reliable if quantum renormalizations are not too strong:
a reasonable criterion to check whether such condition is fullfilled or
not is to require $j_{\rm eff}\gtrsim 0.5$, meaning a quantum
renormalizations of the energy scale of about $50\%$. Such criterion, as
seen in
Fig.~\ref{f.eaj}, is not satisfied for $S=1/2$ and $t\leq 0.5$: 
our $t_{\rm I}(\mu,1/2)$ curve does in fact underestimate the most
anisotropic numerical and experimental data, despite displaying the
correct qualitative behaviour.
A significant improvement may be obtained by using the PQSCHA with the
modified Low Coupling Approximation, as defined in Ref.~\cite{CTVV97},
whose results are shown by the dashed line. This modified version of the
method has the disadvantage, however, of not being capable to cover the
whole $(-1,1)$ $\mu$-interval, because of the loss of solution of its
self-consistent equations when moving towards the isotropic limit (reason 
for the dashed curve in Fig.~\ref{f.tI} to be interrupted).

On the other hand, the almost isotropic region, which is the most
interesting from the experimental point of view, is well described by our
results, even for $S=1/2$, as seen in Fig.~\ref{f.tIing}.
 
\section{Conclusions}
\label{s.concl}

The results found by the PQSCHA for the 2d Heisenberg antiferromagnets
with easy-plane or easy-axis anisotropy, show that quantum fluctuations 
introduce temperature, anisotropy, and spin dependent renormalizations of
both the
energy scale and the anisotropy of the model, consequently
reducing the temperature at which a phase transition, of the BKT
or Ising type, is expected. 
Such reduction, however, is never seen to push the critical
temperature down to zero, so that the whole class of quantum models
described by Eq.~(\ref{e.H}) with $|\mu|\neq|\lambda|$ turns out to be
characterized by a finite temperature transition.

In Fig.\ref{f.tc} a global picture of the situation is visualized by
the $t_c$ curves for different spin values, versus the ratio
$\lambda/\mu$: the central region corresponds to the easy-plane case,
while the two external branches represents the easy-axis ferro- and
antiferromagnetic cases.
The  $t_c$ reduction is seen to be more marked for small spin and in the
ferromagnetic case, thus leading to the conclusion that, 
should $t_c$ vanish for some critical $\nu_c(S)$, we expect $|\nu_c(S)|$ to
be an increasing function of $S$ which is larger in the antiferromagnetic
case, and such that its difference from unity is less than $10^{-2}$. 
In the $S=1/2$ case, despite being the one for which the PQSCHA 
is less precise, our picture completely agrees with that proposed 
by Ding in Ref.~\cite{Ding90} on the basis of quantum MC data.
In the antiferromagnetic $S=1/2$ easy-axis case, on the other hand, 
Branco and De Sousa~\cite{BrancodeS00} predict, from a real space
renormalization
group analysis, the existence of
a critical $\mu_c(1/2)\simeq - 0.8$ at which $t_{\rm I}$ 
should vanish, and detect a reentrance in the $\mu$ dependence 
of such temperature; they found no such anomaly in the ferromagnetic
sector.
This scenario is in contrast not only with ours and Ding's results, but
also with experimental data.

\begin{figure}
\begin{center}
\includegraphics[bbllx=25mm,bblly=30mm,bburx=200mm,bbury=250mm,%
width=70mm,angle=90]{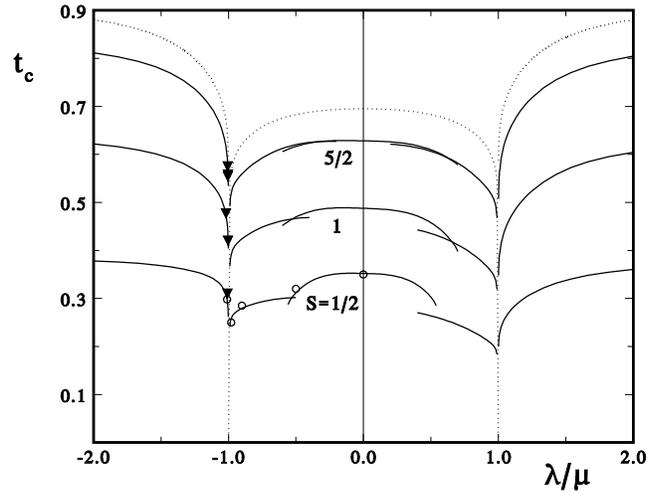}
\end{center}
\caption{Critical temperature {\it vs} $\lambda/\mu$ for the class of
models described by Eq.~\protect\ref{e.H}, with $S=1/2$, $1$, $5/2$, and
$\infty$; symbols and curves as in Figs.\protect\ref{f.tBKT} and
\protect\ref{f.tI}.}
\label{f.tc}
\end{figure}

Finally, from the experimental point of view, our results 
show that anisotropies, even if as small as those measured in most real
compounds, are fundamental ingredients in determining the low-temperature
behaviour of the materials; it is hence essential,
in order to perform a meaningful analysis of the experimental data, to
have at least an estimate of the temperature at which a possible
BKT or Ising-like phase transition should occur. 
Our approach allows a determination of such temperature for any
given value of $S$ and $\nu$, therefore representing an important tool in
the experimental investigation of two-dimensional Heisenberg magnets. 

\section{Acknowledgements}
The authors wish to thank Dr. Luca Capriotti for his contribution in the
early stage of the work. This research has been partially supported by the
COFIN98-MURST fund.
   
%=============  references  =========================================


\begin{thebibliography}{200}

\bibitem{MerminW66}
N.~D. Mermin and H. Wagner, Phys. Rev. Lett. {\bf 17}, 1133 (1966).

\bibitem{Manousakis91}
E. Manousakis, Rev. Mod. Phys. {\bf 63}, 1 (1991).

\bibitem{Berezinskii70}
V.~L. Berezinskii, Zh. Eksp. Teor. Fiz {\bf 59}, 907 (1970).

\bibitem{KosterlitzT73}
J.~M. Kosterlitz and D.~J. Thouless, J. Phys. Condens. Matter {\bf 6},
1181 (1973). 

\bibitem{Onsager44}
L. Onsager, Phys. Rev. {\it 65}, 117 (1944).

\bibitem{ArtsDW90} 
See, e.g., A.~F.~M. Arts and H.~W. de Wijn, in 
{\it Magnetic Properties of Layered Transition Metal Compounds}, 
ed. L.~J. de Jongh, Kluwer Academic Publishers, Dordrecht (1990), pag.
191;
J. Kanamori, in {\it Magnetism}, vol.1, ed. G.T. Rado and H. Suhl,
Academic Press, New York (1963), pag. 127.

\bibitem{CTVV92}
A. Cuccoli, V. Tognetti, P. Verrucchi and R. Vaia, Phys. Rev. A {\bf 45}, 8418
(1992).

\bibitem{CTVVprb92} 
A. Cuccoli, V. Tognetti, P. Verrucchi, and R. Vaia, Phys. Rev. B
{\bf 46}, 11601 (1992).

\bibitem{CTVV97} 
A. Cuccoli, V. Tognetti, P. Verrucchi, and R. Vaia, Phys.
Rev. B {\bf 56}, 14456 (1997). 

\bibitem{CGTVV95}
A. Cuccoli, R. Giachetti, V. Tognetti, R. Vaia, and P. Verrucchi,
J. Phys.: Condens. Matter {\bf 7}, 7891 (1995).

\bibitem{Weyl50}
H.~Weyl in {\it The theory of groups and quantum mechanics}, Dover, New
York (1950).

\bibitem{Villain74} 
J. Villain, J. Phys. (Paris) {\bf 35}, 27 (1974).

\bibitem{HolsteinP40} 
F. Holstein and H. Primakoff, Phys. Rev. {\bf 58}, 1048 (1940).

\bibitem{DysonM56} 
F.J. Dyson, Phys. Rev. {\bf 102}, 1217, (1956); 
S.V. Maleev, Zh. Eskp. Teor. Fiz. {\bf 33}, 1010 (1957).

\bibitem{CTVVprb95}
A. Cuccoli, V. Tognetti, P. Verrucchi, and R. Vaia, 
Phys. Rev. B {\bf 51}, 12840 (1995). 

\bibitem{CCTVV98}
L. Capriotti, A. Cuccoli, V. Tognetti, R. Vaia, and P. Verrucchi,  
Physica D {\bf 119}, 68 (1998).

\bibitem{CTV95}
A. Cuccoli, V. Tognetti, and R. Vaia, Phys. Rev. B {52}, 10221 (1995).

\bibitem{CRTVV00} 
A. Cuccoli, T. Roscilde, V. Tognetti, P. Verrucchi, and R. Vaia,
Phys. Rev. B {\bf 62}, 3771 (2000).

\bibitem{SerenaGL93} 
P~.A. Serena, N. Garc\'\i a, and A. Levanyuk, Phys. Rev. B {\bf 47},
5027 (1993).

\bibitem{GouveaEtal99} 
M.~E. Gouv\^ea,  G.~M. Wysin, S.~A. Leonel, A.~S.~T. Pires, T.~
Kamppeter, and F.~G. Mertens, Phys. Rev. B {\bf 59}, 6229 (1999).

\bibitem{DingM90}
H.-Q. Ding and M.~S. Makivi\'{c}, Phys. Rev. B {\bf 42}, 6827 (1990);
Phys. Rev. B {45}, 491 (1992).

\bibitem{Ding92}
H.-Q. Ding, Phys. Rev. Lett. {\bf 68}, (1992); 
Phys. Rev. B {\bf 45}, 230 (1992).


\bibitem{deWijnWW73} 
H~.W. de Wijn, L.~R. Walker, and R.~E. Walstedt, Phys. Rev. B {\bf 8},
285, (1973).

\bibitem{SkalyoEtal69} 
J. Skalyo, Jr., G. Shirane, R.~J. Birgeneau, and H.~J. Guggenheim,
Phys. Rev. Lett. {\bf 23}, 1394 (1969).

\bibitem{NagataT74} 
K. Nagata and Y. Tomono, J. Phys. Soc. Japan {\bf 36}, 78 (1974).

\bibitem{BirgeneauGS73} 
R~.J. Birgeneau, H~.J. Guggenheim, and G. Shirane, Phys. Rev. B
{\bf 8}, 304 (1973).

\bibitem{SchroederEtal80} 
B. Schr\"oder, V. Wagner, N. Lehner, K.A.M. Kesharwani, and R.
Geick, Phys. Stat. Sol. (b) {\bf 97}, 501 (1980).

\bibitem{CowleyEtal77} 
R.~A. Cowley, G. Shirane, R.~J. Birgeneau, and H.~J. Guggenheim,
Phys. Rev. B {\bf 15}, 4292 (1977).

\bibitem{Ding90} 
H.-Q. Ding, J. Phys.: Condens. Matter {\bf 2}, 7979 (1990).

\bibitem{RossatMignodEtal91} 
J. Rossat-Mignod, L.~P. Regnault, C. Vettier, P. Burlet,
J.Y. Henry, and G. Lapertot, Physica B {\bf 169}, 58 (1991).

\bibitem{BrancodeS00} 
N.~S. Branco and J. Ricardo de Sousa, preprint, cond-mat/0006345 (2000).

\end{thebibliography}
\end{document}